\documentclass[sigconf, 10pt, nonacm]{acmart}
\makeatletter
\renewcommand{\country}[1]{\global\@ACM@countrypresenttrue\ignorespaces}
\makeatother

\usepackage{ifthen, xspace, float}
\usepackage[utf8]{inputenc}
\usepackage{balance}
\usepackage{caption}
\usepackage{subcaption}
\usepackage{soul}
\usepackage{enumitem}
\usepackage{graphicx}
\usepackage{url}
\usepackage{etoolbox, xstring}
\usepackage{booktabs}
\usepackage{array}
\usepackage{pifont}
\usepackage[normalem]{ulem}
\usepackage[english]{babel} 
\usepackage[autostyle]{csquotes}

\DeclareListParser{\doslashlist}{/}
\newcounter{ndnNameComponentCounter}%
\newcommand{\name}[1]{{%
		\setcounter{ndnNameComponentCounter}{0}%
		\renewcommand{\do}[1]{{%
				\ifnumgreater{\value{ndnNameComponentCounter}}{0}{\allowbreak/}{}%
				\ifnumodd{\value{ndnNameComponentCounter}}{}{}%
				\detokenize{##1}}%
			\stepcounter{ndnNameComponentCounter}}%
		``{\fontfamily{cmtt}\small\selectfont\IfBeginWith{#1}{/}{/}{}\doslashlist{#1}}''%
}}

\newboolean{showcomments}
\setboolean{showcomments}{true}
\ifthenelse{\boolean{showcomments}}
{ \newcommand{\mynote}[3]{
    \protect\fbox{\bfseries\sffamily\scriptsize#1}
    {\small$\blacktriangleright$\textsf{\emph{\color{#3}{#2}}}$\blacktriangleleft$}}}
{ \newcommand{\mynote}[3]{}}

\usepackage{xcolor}

\newcommand{\lz}[1]{\mynote{Lixia}{#1}{purple}}
\definecolor{uclablue}{rgb}{0.33, 0.41, 0.58}

\newcommand{\eg}{\textit{e.g.,}\@\xspace}
\newcommand{\ie}{\textit{i.e.,}\@\xspace}

\definecolor{verylightgray}{gray}{0.8}

\AtBeginDocument{%
  \providecommand\BibTeX{{%
    \normalfont B\kern-0.5em{\scshape i\kern-0.25em b}\kern-0.8em\TeX}}}

\usepackage[]{algorithm2e}
\usepackage[export]{adjustbox}

\usepackage{eso-pic,xcolor}
\makeatletter
\AddToShipoutPicture*{%
\setlength{\@tempdimb}{20pt}%
\setlength{\@tempdimc}{\paperheight}%
\setlength{\unitlength}{1pt}%
\put(\strip@pt\@tempdimb,\strip@pt\@tempdimc){%
\makebox(0,-60)[l]{\color{blue}%
NDN Technical Report NDN-0078. \url{http://named-data.net/techreports.html}}
}%
\put(\strip@pt\@tempdimb,\strip@pt\@tempdimc){%
\makebox(0,-85)[l]{\color{blue}%
Version 1: December 2025}
}%
}
\makeatother

\begin{document}
\title{SRM at 30: Lessons from Early Data-Centric Networking and Their Impact on Named Data Networking}

\author{Tianyuan Yu}
\email{tianyuan@cs.ucla.edu}
\affiliation{%
   \institution{UCLA}
   \country{USA}
}

\author{Adam Thieme}
\email{adam@cs.ucla.edu}
\affiliation{%
   \institution{UCLA}
   \country{USA}
}

\author{Junxiao Shi}
\email{junxiao.shi@nist.gov}
\affiliation{%
   \institution{NIST}
   \country{USA}
}

\author{Lan Wang}
\email{lanwang@memphis.edu}
\affiliation{%
   \institution{University of Memphis}
   \country{USA}
}

\author{Lixia Zhang}
\email{lixia@cs.ucla.edu}
\affiliation{%
   \institution{UCLA}
   \country{USA}
}

\begin{abstract}
A 1995 SIGCOMM paper, ``A Reliable Multicast Framework for Light-weight Sessions and Application-Level Framing'', commonly known as SRM, explored a fundamentally new approach to reliable multiparty data delivery. 
Rather than adapting established sender-driven reliable unicast mechanisms to multicast, as most contemporaneous proposals did, SRM introduced a data-centric model in which data receivers recover losses by explicitly requesting missing data.
Thirty years later, we revisit the SRM framework, examining the challenges it faced, the lessons learned, and its influence on the later development of Named Data Networking (NDN). Experimentations with SRM revealed a fundamental \emph{semantic mismatch} between its data-centric framework and IP’s address-based delivery; while the application layer named data, the network layer remained `blind' to those names, resulting in inefficient loss recovery.
NDN resolves this architectural friction by aligning network delivery with the data-retrieval model and by securing data directly rather than securing communication channels.

This retrospective highlights how early insights from SRM informed key design decisions in NDN and illustrates how NDN's design emerged from the cumulative insights gained over decades of networking research and development.

\end{abstract}
\maketitle


\section{Introduction}
The paper ``A Reliable Multicast Framework for Light-weight Sessions and
Application Level Framing''~\cite{SRM95}, published at ACM SIGCOMM 1995, proposed \emph{Scalable Reliable Multicast} (SRM) as a framework for distributed multiparty applications. The paper drew immediate attention from the community. Together with its journal version~\cite{SRM-journal}, which appeared two years later, it has received more than 1,500 citations, reflecting sustained interest in scalable multiparty communication.

The dominant approach at the time was to adapt established unicast transport mechanisms for reliable multicast, such as forming a ring among members~\cite{chang1984reliable}, electing a central controller~\cite{armstrong1992rfc1301}, or establishing multiple point-to-point connections among the members~\cite{birman1993process}. These approaches face significant coordination scaling challenges, especially with dynamic membership changes. 

SRM took a fundamentally different approach. Building on and extending the \emph{Application-Level Framing} (ALF) concept proposed by Clark \cite{SIG90ALF}, SRM assigns each Application Data Unit (ADU) a unique identifier and uses IP multicast to deliver ADUs to all participants in an application group and to collaboratively recover from packet losses. 
The paper demonstrated the effectiveness of the SRM framework by implementing the LBL's shared whiteboard application (wb)~\cite{LBL_wb}, showing that SRM/wb achieved scalability with group size and resilience despite dynamic changes in group membership and network delivery. 

In this paper, we first revisit SRM to review its \emph{data-centered} design features, including application-layer data framing with explicit data naming, receiver-driven collaborative loss recovery, and the eventual delivery of all data to all group members. We refer to these collectively as \emph{data-centric design}, in contrast to IP's address-centric delivery.
We then share two closely related observations from SRM experiments that reveal a common root cause:
(1) conventional protocol mechanisms, such as randomized timers and scope control, are ineffective at bridging the \emph{semantic gap} between SRM's data-centric framework and IP multicast's group address-based delivery model; and
(2) network routers observe only IP multicast group addresses, not the identities of data that SRM seeks to fetch, leading to inefficient loss recovery.
We also note that SRM did not explicitly address security, and that today's unicast, channel-based security model, where protection is applied to the communication channel between endpoints, rather than to the data itself, is ill-suited to collaborative, receiver-driven loss recovery.

Finally, we explain how the design of Named Data Networking (NDN)~\cite{ndn14} incorporated these lessons to develop a data-centric network architecture.
NDN resolves the architectural friction in SRM by making data names, rather than addresses, the primary identifier space for routing and forwarding, and it secures data directly to support receiver-driven data retrieval and loss recovery.

The remainder of the paper proceeds as follows.
Section~\ref{sec:srm} summarizes the technical aspects of SRM, highlights its overlooked contributions, and points out the challenges of efficient receiver-driven loss recovery.
Section~\ref{sec:lesson} investigates the root cause of the inefficiency of SRM's loss-recovery solution, primarily the incompatibility between the framework's data-centric model and address-based network delivery.
Section~\ref{sec:ndn} provides a brief overview of the NDN design.
Section~\ref{sec:eval} analyzes specific scenarios to illustrate how network architectural differences affect loss recovery.
Section~\ref{sec:discuss} addresses a few questions our readers may ask, and Section~\ref{sec:conclude} explores the next step toward data-centric network delivery. 

\section{What Is the Scalable and Reliable Multicast Framework}
\label{sec:srm}
\begin{figure*}[tbh]
    \centering
    \includegraphics[width=\textwidth]{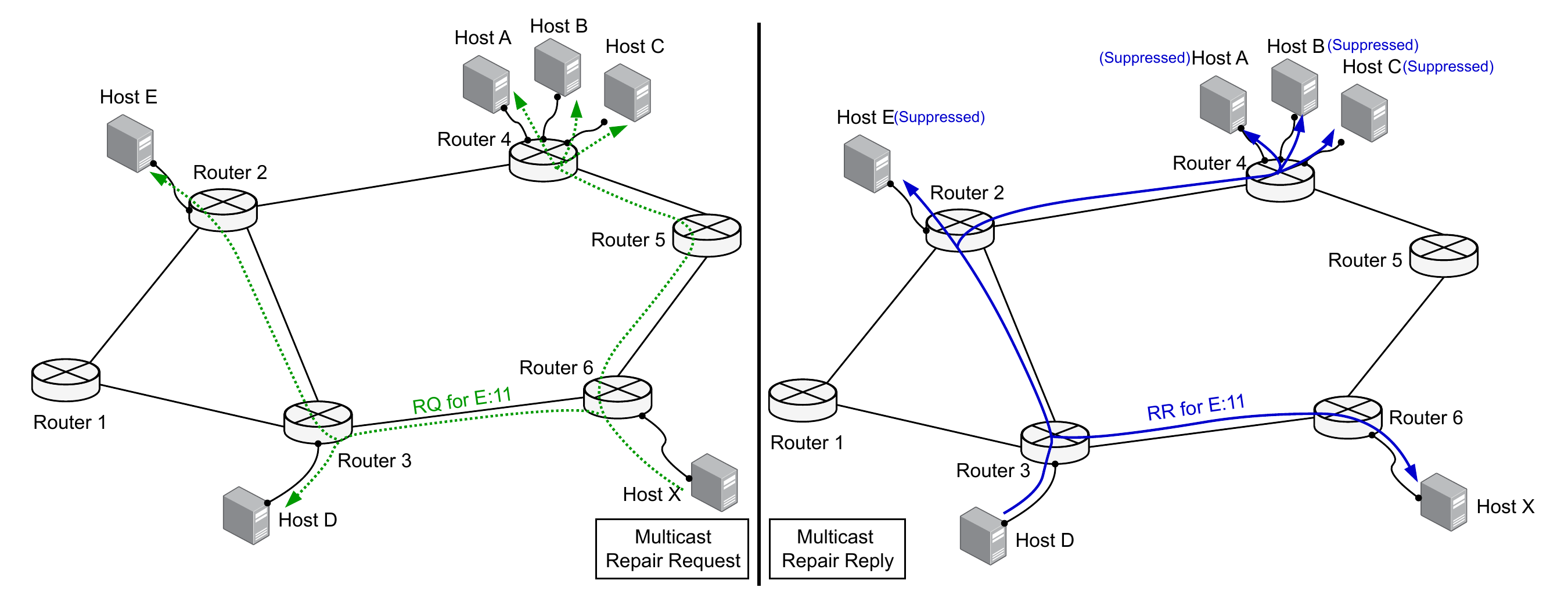}
    \caption{Ideal SRM loss recovery: a single multicast RQ triggers a single multicast RR from the nearest repair node, with all other replies suppressed. Even in this ideal case, both RQ and RR packets are delivered to all group members.}
    \label{fig:srm-topology}
\end{figure*}

Before the development of SRM, efforts on reliable multiparty delivery largely focused on converting the problem to reliable unicast delivery, so they could use the well-established sender-driven, TCP-style reliable unicast transport to achieve the goal~\cite{armstrong1992rfc1301, birman1993process}.  
The SRM paper pointed out that such solutions do not scale well in multicast environments due to heterogeneous receiver conditions and dynamic group membership. 
Instead, SRM proposed \emph{receiver-driven reliability model}, in which each receiver is responsible for its own loss detection and recovery, thereby enabling the solution to scale to any group size and to receivers with heterogeneous data reception states.

This receiver-driven approach is, in turn, enabled by SRM's use of \textbf{\emph{application-level framing}} (ALF) with \textbf{\emph{explicit data naming}}. While ALF had been introduced earlier~\cite{SIG90ALF}, SRM (and the wb application) is the first known design to fully exploit ALF in practice and, critically, to extend it with unique data identifiers. By naming data in application data units (ADUs), SRM eliminated reliance on separate transport protocol state, such as sequence numbers, for reliability, a design that works well in unicast connections but breaks down in multicast due to divergent reception state across different group members.
Assigning unique identifiers to ADUs enables members to explicitly identify missing ADUs, aligning the reliable delivery model with application semantics rather than transport-protocol state.

SRM further reinforced this data-centric model by using IP multicast as the data dissemination substrate for data delivery, loss detection, and loss recovery, eliminating the notion of point-to-point communication.
Rather than addressing packets to specific receivers, each data producer multicasts its data to the group.
To support receiver-based reliability at scale, SRM introduced lightweight mechanisms to inform receivers of new data production and coordinate loss recovery. 

Each group member multicasts session messages periodically to inform the group of its data production state (\ie the sequence number of its most recently produced ADU). If any session message is lost, subsequent messages recover the loss automatically. 
Upon detecting missing data, members multicast Repair Requests (RQ) to the group, making loss recovery a collective responsibility: any member with the requested data may send a Repair Reply (RR) via IP multicast. 
This design exploits the redundancy inherent in multicast groups, reduces recovery latency by avoiding sender bottlenecks, and suppresses both RQ and RR traffic when multiple receivers observe the same loss or multiple members with the requested data attempt to help.

Figure~\ref{fig:srm-topology} shows an example SRM/wb session among members running on hosts $A$, $B$, $C$, $D$, $E$, and $X$. For simplicity, we identify each member by its host name. 
Host $E$ multicasts its ADUs to the group, assigning each a unique identifier of the form \texttt{(E:i)}, where \texttt{i} denotes the ADU's sequence number. 
All other members track the identifiers of the data they receive from $E$.  Suppose host $X$ receives a session message from E indicating \texttt{(E:11)} as $E$'s most recent data identifier, but $X$ has not yet received that ADU. $X$ therefore multicasts an RQ for \texttt{(E:11)}. 
Host $D$, being the nearest neighbor that possesses the data, multicasts an RR carrying \texttt{(E:11)} to the group, thereby suppressing redundant RRs from the other hosts.

In summary, SRM was among the first attempts to explore a \emph{fully realized} data-centric design approach. It moved the focus of data delivery from nodes to data itself.
Instead of sending packets to unicast IP addresses, SRM multicasts everything, including 
(1) application data for efficient delivery, 
(2) session messages to inform each other about the new data productions, and 
(3) RQs and RRs to enable collaborative loss recovery.
This collaborative model eliminates reliance on any single
node and avoids explicit coordination among members.

While SRM's data-centric approach was a breakthrough, implementing it on top of IP multicast introduced significant tension. Because both RQs and RRs are multicast to the entire group, SRM employed randomized timers to prevent multiple members that miss the same data from sending duplicate requests and multiple members that have the data from sending duplicate replies. The SRM paper devoted nearly two-thirds of its pages to investigating mechanisms that suppress duplicate RQs and RRs and encourage group members near the point of packet loss to handle retransmissions. It demonstrated effective solutions, particularly in terms of loss recovery delay.

At the same time, the paper stated that the effectiveness of duplicate suppression by random timers is sensitive to the specifics of network topology and member locations, an observation further confirmed by subsequent experiments~\cite{1999SRM-RecoveryTimer}. 
Even with perfect duplicate suppression for both RQ and RR\footnote{The evaluation results reported in~\cite{SRM-journal} were based on a multicast tree topology in which all nodes are members of the same multicast group, a setting that naturally facilitates duplicate suppression.}, every member of the multicast group still receives two additional packets whenever a single member experiences a loss.  For example, hosts $A$, $B$, $C$, and $E$ in Figure~\ref{fig:srm-topology} all received the RQ from host $D$ and RR from host $X$ even though they did not experience or repair a loss.

\section{Lessons Learned from SRM}
\label{sec:lesson}

The previous section identified two fundamental problems as the causes of inefficiencies in SRM's collaborative loss recovery:
\begin{enumerate}[leftmargin=16pt]
\item The effectiveness of duplicate suppression is highly dependent on network topology and the locations of multicast group members. 
\item Even assuming ideal suppression of both RQs and RRs, a single packet loss by one member still results in two extra packets being delivered to every member of the group.
\end{enumerate}
These problems reflect deeper architectural mismatches rather than shortcomings of specific protocol mechanisms.

The first problem arises because the effectiveness of random timer-based duplication suppression is inherently limited when applied across wide-area networks.
The original expectation of SRM suppression timer design was that, if a member $M$ sends an RQ for a missing ADU \texttt{(E:11)}, the closest member $N$ could immediately respond with an RR carrying the missing \texttt{(E:11)}, thereby suppressing other RQs or RRs.
In practice, however, network delay and topology often prevent timely suppression.
In our example, $N$'s RQ for \texttt{(E:11)} may reach hosts $A$-$D$ at roughly the same time.
Even if host $D$'s random timer expires first and it sends an RR carrying \texttt{(E:11)}, that RR may not reach other hosts before their timers expire, resulting in duplicate RRs carrying \texttt{(E:11)} being delivered to the entire group. 
Similarly, when two distant members miss the same data, there is a high likelihood that they will multicast duplicate RQs. 

The suppression timer setting across wide areas faces an intrinsic dilemma. As Section~\ref{sec:srm} states, the effectiveness of duplicate suppression is sensitive to the specifics of network topology and member locations. Although SRM uses session messages to estimate pairwise round-trip times (RTTs) between members, these estimates are topology-blind. One can improve the effectiveness of duplicate suppression by widening the time window used to select a random value; if the window is set too small, suppression is ineffective because propagation delays dominate; if it is set too large, application-level loss recovery latency increases. 
No single timer configuration can simultaneously satisfy both constraints at scale. 

Other approaches were explored to suppress duplicates, including limiting the scope of RQ multicast delivery by hop count and establishing separate local-recovery multicast groups~\cite{1998LocalErrorSRM, 1013857}. 
The hop-count setting depends on the location of the nearest member with the requested data,
information that is generally unavailable and costly to estimate. Similarly, forming local recovery groups based on shared loss patterns increases system complexity and introduces additional control overhead. These techniques mitigate symptoms but do not address the root cause of inefficiency.

The second problem arises because multicasting to the entire group is SRM's only mechanism for signaling loss and delivering repairs.
A member $M$ must multicast its RQ to inform others of its missing data, even though not all group members need to know it. Likewise, RRs are multicast to the entire group, even though not all group members need the retransmission.
This inefficiency becomes particularly pronounced when a single member, say host $X$ in Figure~\ref{fig:srm-topology}, is connected to Router 6 via a lossy link and produces a high volume of RQs. 

Ideally, host $X$ could direct its RQ to the nearest member that holds the missing data, such as host $D$. However, SRM deliberately eliminates point-to-point communication and provides no mechanism for identifying which member has the data or is topologically closest. 
More fundamentally, neither end hosts nor routers know where data copies reside. The network can observe only multicast group addresses, not the identities of data items or their locations.

In summary, there is no effective way within SRM to eliminate either inefficiency. The fundamental reason is that the multicast group is the basic unit of packet delivery, while
loss recovery operates on named data items. Multicasting RQs and RRs therefore necessarily delivers recovery traffic to group members that neither need nor benefit from it. 
This mismatch between data-centric recovery semantics and IP multicast's group address-based packet delivery cannot be bridged by protocol mechanisms such as random delays or scoped delivery.

We observe that an ideal solution would allow routers to recognize the semantic intent of RQs -- namely, fetching a specific named data item -- and to forward requests toward the nearest available copy. Routers equipped with this knowledge could cache data and satisfy recovery requests locally, preventing unnecessary propagation. Making data identifiers visible to the network layer would therefore enable efficient, localized recovery.


Unfortunately, although SRM assigns names to individual ADUs, these names are not visible to the network. At the time, IP multicast groups were viewed as semantic application constructs decoupled from topology, and their receiver-driven creation appeared to align with SRM's receiver-driven recovery. Consequently, SRM was designed under the assumption that no special support from the underlying IP network would be required. This design choice ultimately exposed a fundamental incompatibility between SRM's data-centric framework and IP's address-based delivery model.

Finally, given its historical context, SRM did not explicitly address communication security. Today, secure communication is mandatory, yet the prevailing channel-based security mechanisms are poorly suited to collaborative, receiver-driven recovery. 
Advancing data-centric networking thus requires a corresponding shift toward data-centric security.

Taken together, these observations show that SRM’s loss-recovery inefficiencies arise from a fundamental architectural mismatch between data-centric recovery semantics and IP’s address-based delivery model, a lesson that we next place in a broader context by examining how modern systems attempt to compensate for this mismatch.

\section{NDN: A Data-Centric Framework}
\label{sec:ndn}

\begin{figure*}[ht]
    \centering
    \includegraphics[width=\textwidth]{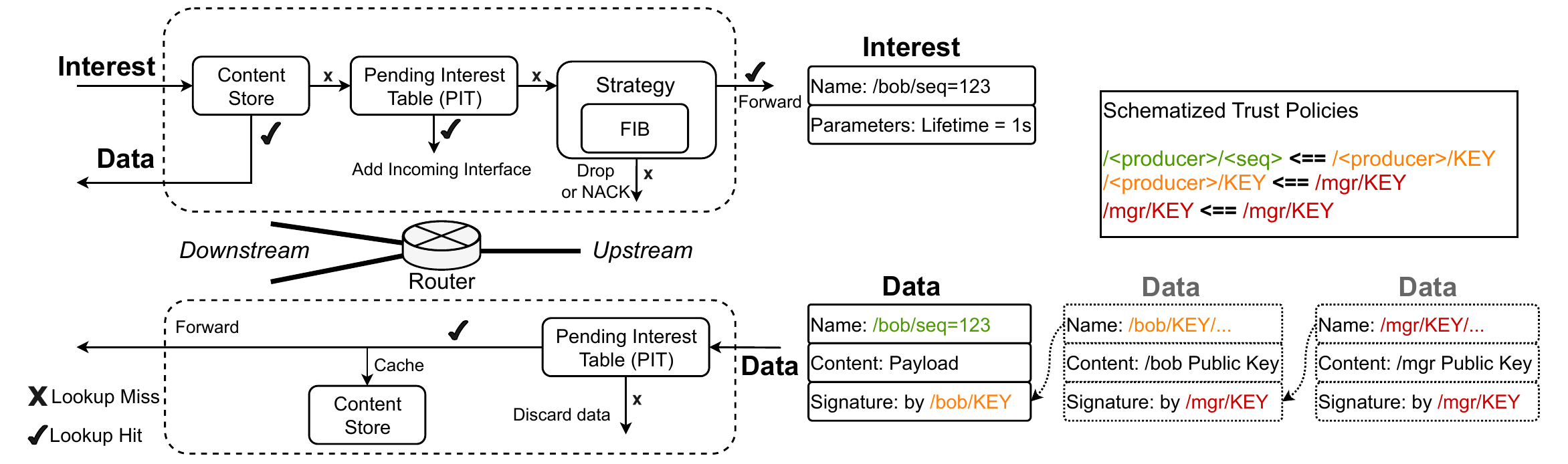}
    \caption{Forwarding pipelines at an NDN Router: upstream indicates the direction of the data producer and downstream is towards the data consumer. The consumer expresses Interest to network to fetch Data that are validated by schematized trust policies.}
    \label{fig:ndn101}
\end{figure*}

The development of NDN began with its predecessor, \emph{Content-Centric Networking} (CCN), as envisioned by Jacobson in his 2006 Google Talk~\cite{NewWayToLookAtNetworking2006}. The design was renamed \emph{Named Data Networking} when it was funded by the NSF in 2010 as one of the future Internet architecture design efforts~\cite{NSF_FIA_2010}.
The design of NDN drew on decades of Internet research and development, including datagram delivery, the end-to-end principle, IP multicast, ALF, and, in particular, the lessons from the SRM experiments.

NDN advanced the data-centric model started by SRM with three fundamental differences:
\begin{enumerate}[leftmargin=16pt]
\item Adding cryptographic protection to all data. Data is not only semantically named but also encrypted and signed during production.  
\item Instead of multicasting data, producers publish named, secured data items by making them available for retrieval.  Consequently, data consumers (members in a group) fetch all data by name, not just the missing pieces.
\item The data-centric model is extended to the network layer, where network nodes forward data requests based on data names and return matching data to requesters.
\end{enumerate}
Together, these changes align data dissemination, loss recovery, and security under a single, name-based abstraction.

Adding security protection to data directly represents a new direction in securing networked systems. 
Similarly, extending the data-centric model to the network layer opens a solution space in which several persistent IP network problems either disappear or become easier to address.
The remainder of this section briefly describes NDN's core concepts: Interest-Data exchange, the NDN security framework, data-centric networking by stateful forwarding, support for reliable data retrieval, and the deployment of NDN as an overlay.

\subsection{Named, Secured Data as the Basic Building Block}
A network architecture begins by choosing its basic building block. IP uses datagrams as the basic unit of delivery,
forwarding individual packets to destination addresses~\cite{IP-design1988}.
NDN uses \emph{named, secured data packets} as its basic building block~\cite{2010NDNProject}, and supports data retrieval by name.
Figure~\ref{fig:ndn101} illustrates the basic operation of NDN data retrieval.

\emph{Data producers} publish their data in NDN \emph{Data packets}, each identified by a semantic name (derived from the DNS namespace) and carrying a cryptographic signature that binds the name to the content; the content may also be encrypted as needed.
Data producers announce name prefixes to routing protocols to make their data reachable.

\emph{Data consumers} send \emph{Interest packets} carying the names of the desired data.
Network routers forward Interests toward data producers according to routing information; each router along the way  records where Interests arrive from. 
When an Interest encounters a matching data packet, either at the original producer or a router cache, the data packet is returned by following the reverse path of the Interest, as we explain in Section~\ref{sec:forward} with more details.
Consumers ensure retrieval reliability by retransmitting Interests when data is not returned within an expected time window.

\subsection{Adding Security Protection on Data}
\label{sec:security}
NDN integrates cryptographic protection directly into its architecture by securing every data item during production and requiring consumers to verify all received data. 
To produce and consume named, secured data, each networked entity, be it a producer or consumer or both, must possess a set of security parameters, including certificate, trust anchors, and a set of security policies defined by the system controller and applications and encoded as trust schemas. 
Security policies can be expressed as relations between cryptographic key names and data names, specifying which keys are authorized to sign which data~\cite{yu2015schematizing}. 
 
Each entity undergoes a \textit{bootstrapping} process to obtain its initial set of security parameters~\cite{2021plug-play}. Certificates and trust schemas are themselves named,
secured data and can be retrieved and updated dynamically.
This design allows security state to evolve continuously
without introducing external configuration channels.

As illustrated in Figure~\ref{fig:ndn101}, the signature section of a Data packet includes a key locator that names the signer's certificate, forming a verification chain that terminates at a trust anchor.
As described earlier, the signing relation is governed by security rules defined by systems and applications, expressed as relations between data names and signer names and encoded as a trust schema. Each key can sign data only within its permitted namescope.
In Figure~\ref{fig:ndn101}, the trust schema defines that \name{/Bob/seq=123} can only be signed by the producer with the same name, only the entity \name{/mgr} can certify the data producers, and the self-signed certificate of \name{/mgr} is the root of trust for \name{/Bob}.

Consumers validate each Data packet by following the above trust chain and enforcing trust-schema rules that bind data names to authorized signers.


\subsection{Network Layer: Routing and Stateful Forwarding}
\label{sec:forward}

To deliver packets over a network, the network must provide routing and forwarding as fundamental functional requirements.
A network routing protocol sets up the forwarding information base (FIB) at each router, which forwards packets according to the FIB.
Routing is independent of the type of namespace used for packet delivery (e.g., IP uses IP address space, and NDN uses semantic namespaces); thus, an NDN network also needs to run a routing protocol to populate router FIBs, and multiple NDN routing protocols have been developed~\cite{2018NLSR, 2024NDN-dv}. 

NDN Interest packets carry the names of requested data but not the requesters' identities. To return data to the requesters, the network must maintain traces of forwarded Interest packets. NDN employs a stateful forwarding plane that maintains per-Interest state for all Interests that have been forwarded but for which the data has not yet returned, so that the data can follow the reverse paths of the Interests back to the requesters. 
This per-Interest state creates a feedback loop, enabling NDN routers to detect and recover from failures at a round-trip-time scale.

When a router $R$ receives an Interest $I$, $R$ first checks whether its Pending Interest Table (PIT) already contains an entry for the name carried in $I$.
\begin{itemize}[leftmargin=12pt]
\item If yes, $R$ simply adds $I$'s incoming interface to the entry but does not forward it (Interest aggregation).
\item If not, $R$ records $I$ in PIT and forwards $I$ according to the FIB and the forwarding strategy for $I$'s namespace\footnote{NDN allows a network to support various forwarding strategies, such as shortest path only or multipath forwarding, defined for specific namespaces. The forwarding strategies can be propagated through routing protocols. We omit the details here.}.
\end{itemize}
Once $I$ reaches the matching Data $D$, $D$ is forwarded back to its requester(s). Each router along the path forwards $D$ to all incoming interfaces of the Interests in the corresponding PIT entry, then stores $D$ in its local cache (called the Content Store, CS) before moving the PIT entry for $I$. Upon receiving data $D$, each consumer verifies the authenticity and integrity of $D$ according to its local trust schema.
In addition to the Content Store at routers, which serves as opportunistic in-network storage, NDN also supports data repositories (Repo, in short), which provide managed in-network storage to keep data available to support asynchronous communications among group members.

The procedure above demonstrates that NDN has built-in multicast Data delivery.
By aggregating Interests with the same data name on their paths toward the producer, PIT entries at NDN routers along the path form a multicast tree rooted at the producer and branching to all requesting consumers, enabling the requested data to flow down the tree.


\subsection{Reliable Data Retrieval}
\label{subsec:sync}
NDN places control over reliability at the data consumer.
Consumers detect missing data through application logic
and recover losses by retransmitting Interests. This design
avoids imposing a one-size-fits-all reliability mechanism
and naturally accommodates heterogeneous reception states.

To fetch data by name, consumers need to know the names of available data.
NDN meets this need through its transport protocol, dataset-state-sync (Sync)~\cite{syncsok}.
After exploring a wide range of the Sync design space, the latest Sync protocol, State Vector Sync (SVS)~\cite{tr-svs}, provides simple and resilient synchronization of dataset states among all group members.
SVS encodes the data production state of all members in an application group into a state vector, where each element in the vector is a tuple containing a producer name and its sequence number.
The state vector is encoded as a named, secured data item, which is carried in an NDN Interest packet, called Sync Interest\footnote{In NDN, Interest packets retrieve Data packet; we call such Interest \emph{data interest}. However, protocol design is an engineering practice in which decisions are made through trade-offs. NDN allows an Interest packet to carry data in its ``application parameter'' field under a few exceptional circumstances. Sync is one of them. The detailed reasoning will be elaborated in a future publication.}.
Each member in a Sync group multicasts its Sync Interest whenever it produces new data items, or periodically as a soft-state protocol. In the absence of new data, SVS's suppression timer ensures that the group generates approximately one Sync Interest per refresh period, rather than every member sending one. 

Consumers can detect missing data from received Sync Interests and send data Interests to fetch it. If more than one consumer misses the same data, their data Interests with the same name are aggregated by routers and can be satisfied from a router cache before reaching data producers.

NDN's combination of named data, Interest aggregation,
in-network caching, and receiver-driven retransmission
enables scalable and efficient reliable data delivery.

\subsection{Rolling out NDN}
\label{sec:ndnoverlay}
NDN is a full-stack network architecture, encompassing naming, data-centric security, synchronization as transport semantics, and network-layer forwarding. Its deployment begins with NDN-based applications that generate and consume named data packets, which in turn require a network substrate to forward these packets.

Within local environments where ad hoc connectivity is available, NDN entities can directly exchange Interest and Data packets over existing link layers such as Ethernet, Wi-Fi, or Bluetooth, forming small, self-contained NDN islands. To interconnect these islands across wide areas, the NDN Testbed~\cite{ndn-testbed} has been deployed and operational for more than a decade as a global-scale overlay network. In this deployment, NDN routers are interconnected via configured tunnels over existing point-to-point transports (\eg TCP and UDP). Existing NDN software libraries further simplify deployment by enabling applications to automatically establish connections to testbed routers using TCP, UDP, QUIC, or WebSocket tunnels.

Although the NDN Testbed comprises only on the order of twenty routers, these nodes are distributed across four continents, providing wide-area connectivity and a realistic operational environment for experimenting with data-centric networking at Internet scale, without requiring any modifications to the underlying IP substrate.

Note that IP itself was initially deployed as an overlay atop legacy telecommunication infrastructure, with its widespread adoption driven by strong demand from the user community for IP-based applications. NDN is following a similar path, with deployment likewise pulled forward by application needs rather than by changes to the existing network infrastructure.


\begin{figure*}[ht]
    \centering
    \includegraphics[width=\textwidth]{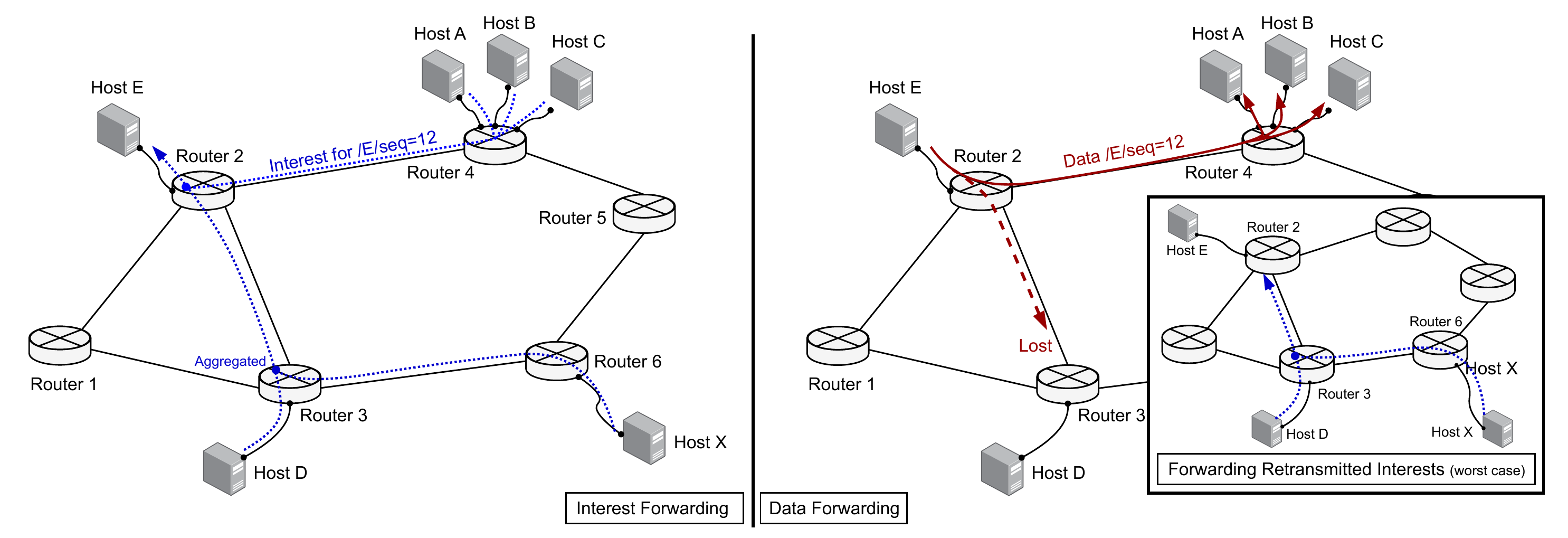}
    \caption{Interest and Data forwarding in NDN under packet loss. The name \texttt{/E/seq=12} refers to the \texttt{(E:12)}. (left)  Interest forwarding and aggregation by name construct a transient multicast tree rooted at router~2. (right) Data forwarding follows the reverse paths of aggregated Interests; packet loss on the 2--3 link affects only the downstream receivers. (inset) Upon timeout, retransmitted Interests from the loss neighborhood are aggregated and satisfied without affecting other members, localizing recovery traffic.}
    \label{fig:compare}
\end{figure*}

\section{Comparison of SRM and NDN Loss Recovery via an Example}
\label{sec:eval}
In this section, we demonstrate, using a concrete example, how aligning network delivery with the data-centric framework
of higher layers eliminates the inefficiencies observed in SRM’s loss recovery. We use the same network topology and data
distribution scenario shown in Figure~\ref{fig:srm-topology}, assuming that all nodes in the figure run the NDN protocol for data delivery.

When host~E produces a new data item \texttt{(E:12)}, it multicasts a Sync Interest carrying its latest dataset state vector 
\[
[A\!:\!123, B\!:\!223, C\!:\!15, D\!:\!941, \textbf{E\! : \!12}, X\!:\!431],
\]
Upon receiving E's Sync Interest, other group members send data Interest for \texttt{(E:12)}.
As illustrated in Figure~\ref{fig:compare}, each router aggregates Interests that carry the same data name and forwards only one upstream toward the producer, host~E.
For example, router~4 receives three Interests for \texttt{(E:12)} from hosts A--C, it forwards the first to router~2 and records the incoming interfaces for the remaining two.
Similarly, router~3 receives Interests for data \texttt{(E:12)} from host~D and router~6, forwarding only one toward host~E. 

As a result of Interest aggregation, Host~E receives a single Interest and responds with a Data packet carrying \texttt{(E:12)}. Router~2 caches the data and forwards it downstream toward Routers 3 \& 4, which in turn forward it further downstream towards all requesters. This process constructs a transient multicast delivery tree rooted at the data producer, without requiring explicit multicast signaling.

NDN provides a unified mechanism for retrieving both newly
produced and previously lost data through the same
Interest--Data exchange.
When data \texttt{(E:12)} is lost over the link between routers 2 and 3, the retransmission timers at
hosts D and X expire, causing them to resend Interests for \texttt{(E:12)}.
Regardless of which retransmission occurs first,
the retransmitted Interest reaches router 2, which satisfies
the Interest from its cache and forwards the Data back toward
the affected receivers via router 3.

In contrast to SRM’s multicast-based Repair Requests and
Repair Replies, retransmitted NDN Interests are forwarded
toward the data and can be satisfied by router caches.
Recovery traffic is therefore confined to the loss region
rather than being propagated to the entire group. This
behavior follows directly from NDN routers’ ability to
distinguish Interests from Data, forward requests by name,
and cache passing data packets.
 
By contrast, although SRM also supports receiver-driven loss recovery, its only available mechanism is multicasting all traffic, since the underlying IP delivery model recognizes only unicast and multicast addresses, not data identifiers. 
As a result, SRM cannot localize recovery traffic, even when the location of the nearest data copy may be
topologically close to a requester. 

In summary, both SRM and NDN support reliable multiparty communication.
A fundamental distinction between two-party and multiparty communication is that group members may be in different data-reception states at any given time.
Both systems address this challenge using receiver-driven loss recovery, where receivers request missing \emph{data} without targeting any specific node.
More fundamentally, reliable multiparty communication inherently requires a data-centric framework: recovery operates on named pieces of data (what is missing), not on host addresses (where it might be). Only NDN enables the network to forward recovery requests toward the data itself, allowing efficient, localized recovery. SRM cannot achieve this behavior because its underlying network recognizes only IP addresses rather than named data.

\section{Discussion}
\label{sec:discuss}



Although the SRM paper was published 30 years ago, its goal of providing scalable, reliable multicast for multiparty applications remains largely unmet today. During the early days of IP multicast development in the late 1980s, David Clark famously remarked, “If you think you solved a problem, say the word multicast.” At the time, this observation captured a widely shared intuition about the difficulty of multicast, without a clear articulation of its root causes. 
The analysis in Section 3 allows us to revisit this statement with the benefit of hindsight. SRM’s experience reveals that the fundamental challenge lies not in protocol mechanics, but in the mismatch between data-centric communication semantics and an address-centric network delivery model.
In this section, we place this lesson in a broader context by examining how modern systems attempt to work around this mismatch and what remains missing in today’s Internet architecture.
\vspace*{6pt}

\noindent\textbf{\emph{Lessons from CDN Overlays:}}\quad
One practical response to the mismatch between data-centric application semantics and address-based network delivery has been the deployment of \emph{Content Distribution Networks} (CDNs), which approximate data-centric delivery through application-layer overlays without changing the underlying IP substrate.

CDN nodes are explicitly provisioned and interconnected by providers to fetch, cache, and serve content identified by URLs, forwarding requests within the overlay when local copies are unavailable. Through this process, CDNs deliver content by name (URL) and demonstrate the practical benefits of data-centric techniques. 

However, these benefits do not generalize to the Internet as a whole. 
CDNs are specialized infrastructures operated by a small number of providers, serve a limited set of paying content sources, and do not offer general-purpose data-centric delivery for arbitrary application traffic. As a result, multiparty applications continue to bridge the gap between data-centric communication needs and point-to-point network delivery through application-specific overlays and middleware, rather than relying on a shared data-centric architectural substrate.
\vspace*{6pt}

\noindent\textbf{\emph{The Implementation of Today's Multiparty Apps:}}\quad
Many modern multiparty applications, including group messaging and collaborative systems, exhibit inherently data-centric semantics: participants produce and consume shared data objects, often asynchronously and with heterogeneous reception states. 

In practice, these applications are typically implemented using centralized brokers or full-mesh peer-to-peer connections. Full-mesh designs do not scale to Internet-scale deployments. In contrast, broker-based architectures introduce application-specific overlays that reimplement data dissemination, replication, and loss recovery on top of point-to-point transports. 
This repeated reinvention across applications highlights the absence of a general-purpose data-centric communication substrate in today’s Internet architecture.
\vspace*{6pt}

\noindent\textbf{\emph{``HTTP as the Narrow Waist of the Future Internet''}}\quad
This 2010 proposal~\cite{popa2010http} argued that HTTP has become the de facto narrow waist of the Internet, therefore its dominance enables content-centric functionality through named resources and extensible intermediaries, without replacing IP.
While HTTP successfully names resources, it was not designed as a general-purpose communication substrate. 
Its client–server, request–response model lacks native support for forwarding requests toward data, for group communication semantics, for receiver-driven repair, or for coordinated data dissemination. 
Supporting multiparty communication over HTTP therefore requires pushing loss recovery, replication, and state management into application-level middleware. 

In contrast, NDN aligns network delivery directly with data-centric communication semantics, allowing multiparty communication to be expressed as data retrieval without relying on application-specific indirection.
\vspace*{6pt}

\noindent\textbf{\emph{The Need for Multicast Routing Support}}\quad
Although NDN avoids SRM's inefficiency from multicasting everything 
by converting multicast repair signaling into data retrieval by name, NDN still requires network multicast routing support for Sync Interest multicast, which serves to share publication state among group members. 
Sync Interests can be viewed as an evolution of SRM session messages: rather than advertising a single producer’s sequence number, a Sync Interest encodes the publication state of all producers in the group. 
However, this does not change the fundamental nature shared between the two: both require network multicast routing support to disseminate the group publication state, a point we explore further in forthcoming work~\cite{NDN-mcast}. 

\section{Looking Forward}
\label{sec:conclude}

As we noted in Section~\ref{sec:lesson}, SRM's design decision of multicasting everything was largely driven by the constraints of the underlying address-based IP delivery
service. 
This paper argues that scalable, reliable multiparty
communication requires aligning network delivery with
data-centric application semantics.
However, the same network delivery model constraint remains.
 
In this section, we illustrate how one can experiment with multiparty applications using data-centric delivery over an NDN overlay.
As noted in Section~\ref{sec:ndnoverlay}, NDN entities can communicate over any available connectivity, whether physical or virtual.  This flexibility enables incremental
deployment without requiring changes to the underlying
IP network.

The NDN Testbed~\cite{ndn-testbed} has been deployed and operational for more than a decade as a multicontinental overlay network. It serves as an operational platform that natively supports data-centric communication semantics, and multiple applications have been running continuously over the NDN overlay.

One such application is Docker Registry over NDN (DRON)\footnote{\url{https://github.com/yoursunny/DRON}}. 
In the current Docker Registry API~\cite{Docker-Registry-HTTP-API-V2}, blob retrieval (\ie pulling image layers) accounts for the majority of network traffic. DRON places a producer co-located with the Docker registry and a consumer at the client host. 
The consumer receives Docker API requests from the local Docker Engine, translates blob retrieval requests into NDN object-retrieval Interests, and proxies all other requests to the registry over HTTPS. 
The producer translates incoming Interests back into Docker blob retrieval requests and returns the retrieved blobs to the network. 
This design allows each blob to be fetched from the registry only once, after which the NDN Testbed efficiently distributes the data to all requesting clients.

Another application running on the NDN overlay is Ownly\footnote{\url{https://ownly.work}}, a decentralized collaborative editor implemented as an in-browser application. Users interact with local Ownly clients to create shared workspaces and collaborate on group files, with each participant maintaining a local copy. 
Clients encrypt and sign locally-generated updates and communicate directly with one another, with NDN Testbed routers acting solely as packet forwarders. Each update is published as newly named, secured data, and SVS is used to propagate the dataset state changes (see Section~\ref{sec:ndn}). Upon detecting new or missing data items, clients retrieve, validate, decrypt, and apply them locally. 
NDN’s data-centric delivery semantics ensure that each update is transmitted only once, even as group size grows.

Together, these deployments demonstrate that NDN is not
merely an architectural proposal but a viable platform for
building and evaluating multiparty applications today.
The availability of a long-running testbed offers an immediate opportunity for developers to explore data-centric designs in practice, whether by running applications directly over the existing NDN overlay, deploying their own NDN nodes to interconnect with it, or constructing one's own NDN overlay to experiment with new multiparty communication patterns.
%

\bibliographystyle{ACM-Reference-Format}
\balance
\bibliography{refs}

\end{document}